\documentclass[preprint]{aastex}

\shorttitle{CHEOPS performance for exomoons}
\shortauthors{Simon et al.}

\begin{document}


\title{CHEOPS performance for exomoons: \\
    The detectability of exomoons by using optimal decision algorithm}


\author{A. E. Simon\altaffilmark{1,2}}
\affil{Physikalisches Institut, Center for Space and Habitability, University of Berne, CH-3012 Bern, Sidlerstrasse 5}
\email{attila.simon@space.unibe.ch}

\author{Gy. M. Szab\'o\altaffilmark{1}}
\affil{Gothard Astrophysical Observatory and Multidisciplinary Research Center of Lor\'and E\"otv\"os University, H-9700 Szombathely, Szent Imre herceg u. 112.}
\email{szgy@gothard.hu}

\author{L. L. Kiss\altaffilmark{3}}
\affil{Konkoly Observatory, Research Centre for Astronomy and Earth Sciences, Hungarian Academy of Sciences, H-1121 Budapest, Konkoly Thege. Mikl\'os. \'ut 15-17.}
\email{kiss@konkoly.hu}

\author{A. Fortier}
\affil{Physikalisches Institut, Center for Space and Habitability, University of Berne, CH-3012 Bern, Sidlerstrasse 5}
\email{andrea.fortier@space.unibe.ch}

\and

\author{W. Benz}
\affil{Physikalisches Institut, Center for Space and Habitability, University of Berne, CH-3012 Bern, Sidlerstrasse 5}
\email{willy.benz@space.unibe.ch}


\altaffiltext{1}{Konkoly Observatory, Research Centre for Astronomy and Earth Sciences, Hungarian Academy of Sciences, H-1121 Budapest, Konkoly Thege. Mikl\'os. \'ut 15-17.}
\altaffiltext{2}{Gothard Astrophysical Observatory and Multidisciplinary Research Center of Lor\'and E\"otv\"os University, H-9700 Szombathely, Szent Imre herceg u. 112.}
\altaffiltext{3}{Sydney Institute for Astronomy, School of Physics, University of Sydney, NSW 2006, Australia}


\begin{abstract}
Many attempts have already been made for detecting exomoons around transiting exoplanets but the first confirmed discovery is still pending. The experience that have been gathered so far allow us to better optimize future space telescopes for this challenge, already during the development phase. In this paper we focus on the forthcoming {\it CHaraterising ExOPlanet Satellite (CHEOPS)},describing an optimized decision algorithm with step-by-step evaluation, and calculating the number of required transits for an exomoon detection for various planet-moon configurations that can be observable by CHEOPS. We explore the most efficient way for such an observation which minimizes the cost in observing time. Our study is based on PTV observations (photocentric transit timing variation, Szab\'o et al. 2006) in simulated CHEOPS data, but the recipe does not depend on the actual detection method, and it can be substituted with e.g. the photodynamical method for later applications. Using the current state-of-the-art level simulation of CHEOPS data we analyzed transit observation sets for different star-planet-moon configurations and performed a bootstrap analysis to determine their detection statistics. We have found that the detection limit is around an Earth-sized moon. In the case of favorable spatial configurations, systems with at least such a large moon and with at least Neptune-sized planet, 80\% detection chance requires at least 5-6 transit observations on average. There is also non-zero chance in the case of smaller moons, but the detection statistics deteriorates rapidly, while the necessary transit measurements increase fast. After the CoRoT and Kepler spacecrafts, CHEOPS will be the next dedicated space telescope that will observe exoplanetary transits and characterize systems with known Doppler-planets. Although it has smaller aperture than Kepler (the ratio of the mirror diameters is about 1/3) and is mounted with a CCD that is similar to Kepler's, it will observe brighter stars and operate with larger sampling rate, therefore the detection limit for an exomoon can be the same as or better, which will make CHEOPS, a competitive instruments in the quest for exomoons.
\end{abstract}





\section{Introduction}

During the past twenty years, astronomers have discovered a large set of exoplanets, which has sparked an excitement in the community whether these planets may host a detectable and/or a habitable exomoon. The number of confirmed exoplanets exceeds 1000 but so far there is not a single case where the existence of an exomoon could have been demonstrated. Here we have to note, that although Bennett et al. (2014) detected a microlensing event, which was interpreted as signal from an exomoon orbiting a gas giant, the interpretation of the result is doubtful.

Sartoretti \& Schneider (1999) argued that a moon around an exoplanet can cause a measurable timing effect in the motion of the planet and therefore its transit timing (known as TTV). Since then many efforts have been made to develop and test such algorithms that can help us answer the question whether an exomoon can be discovered around exoplanets. The series of dedicated studies started with the photocentric transit timing variation (PTV, TTV{\footnotesize p} in Szab\'o et al. 2006, Simon et al. 2007), followed by the detection of mutual transit events between planet and its moon(s) (Cabrera \& Schneider 2007), the transit duration variation (TDV; Kipping 2009b), the influence of lunar-like satellites on the infrared orbital light curves (Moskovitz et al. 2009), direct photometric detection of moons and rings (Tusnski \& Valio 2011). Recent studies include the scatter analysis of phase-folded light curves (Simon et al. 2012), the HEK Project (direct detection of photometric light curve distortions and measurements of timing variation like TTV and TDV, Kipping et al. 2012, 2014) and the analysis of orbital sampling effect (Heller 2014). In addition, there has been a plethora of other techniques identified that can play important role in confirming such detections. These include the Rossiter-McLaughlin effect (Simon et al. 2009, 2010; Zhuang et al. 2012), timing variations in pulsars' signal (Lewis et al. 2008), microlensing (Han \& Han 2002), excess emission due to a moon in the spectra of distant jupiters (Williams \& Knacke 2004), direct imaging of tidally heated exomoons (Peters \& Turner 2013), plasma tori around giant planets by volcanically active moons (Ben-Jaffel \& Ballester 2014) and modulation of planetary radio emissions (Noyola et al. 2014). 
 
Also, there were attempts to identify the potential Kepler candidates that could host a possible exomoon (Szab\'o et al. 2013; Kipping et al. 2012, 2014, the HEK Project). Even though Kepler is theoretically capable of such a detection, there is no compelling evidence for an exomoon around the KOI (Kepler Objects of Interest) targets so far.

In this paper we turn to the next space telescope, CHEOPS, and characterize chances of an exomoon detection by using the method of the photocentric transit timing variation (PTV, TTV{\footnotesize p} Szab\'o et al. 2006, Simon et al. 2007). Our aim is to explore the capabilities of CHEOPS in this field and calculate how many transit observations are required to a firm detection in the case of a specific range of exoplanet-exomoon systems. We did this according to the CHEOPS science goals and planets to be observed, without investigating the theoretically expected occurrence rate and probability of exomoons generally. We test a decision algorithm, which gives the efficiency of discovering exomoons without misspending the expensive observing time. We perform simulations and bootstrap analysis to demonstrate the operation of our newly developed decision algorithm and calculate detection statistics for different planet-moon configurations. Finally, we introduce a folding technique modified by the PTV to increase the possibility of an exomoon detection directly in the phase-folded light curve, which is similar to the idea presented by Heller (2014).

\section{Transit Timing Variations: the meaning of TTV and PTV}

First, for the sake of clarity, we have to note that the photocentric transit timing variation (PTV) differs from the conventional transit timing variation (TTV). Because of their blurred meaning (PTV and TTV were called as TTV{\footnotesize p} and TTV{\footnotesize b}, respectively by Simon et al. 2007) we clarify briefly the differences and show their usabilities and limitations in the following.

\begin{figure}
\centering
\plotone{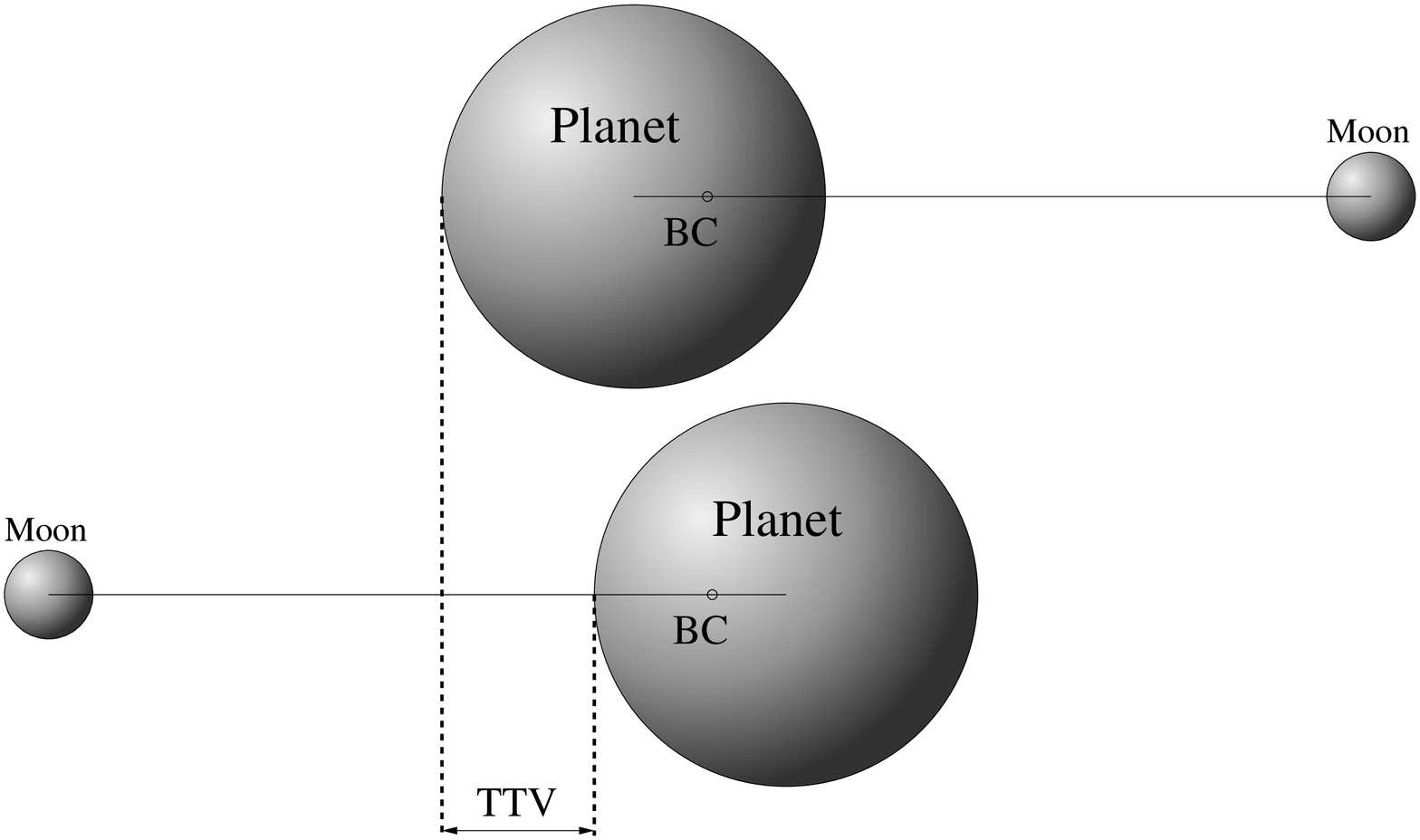}
\caption{The conventional TTV: the time shift due to the revolution of the companion around the common barycenter (Sartoretti \& Schneider 1999). Not to scale.}
\label{ttvb}
\end{figure}

Sartoretti \& Schneider (1999) suggested that a moon around an exoplanet can be detected by measuring the variation in the transit time of the planet due to gravitational effects. In their model, the barycenter of the system orbits the star with a constant velocity, and transits strictly periodically. As the planet revolves around the planet-moon barycenter, its relative position to the barycenter is varying, so the transit of the planet starts sometimes earlier, sometimes later (Fig. \ref{ttvb}). This time shift is the conventional TTV:

\begin{equation}
TTV \sim \frac{m_s}{m_s+m_p} \approx \frac{m_s}{m_p} = \chi\vartheta^3,
\end{equation}
where $m_s$ and $m_p$ are the masses of the moon and the planet, $\chi$ and $\vartheta$ are the ratios (moon/planet) of the densities and radii, respectively. We note that TTV can be caused not only by an exomoon itself. Several other processes, for example an additional planet in the system (e.g.: Agol \& Steffen 2005, Nesvorný et al. 2014), exotrojans (Ford \& Gaudi 2006), periastron precession (P\'al \& Kocsis, 2008) can also cause TTVs.

\begin{figure}
\centering
\plotone{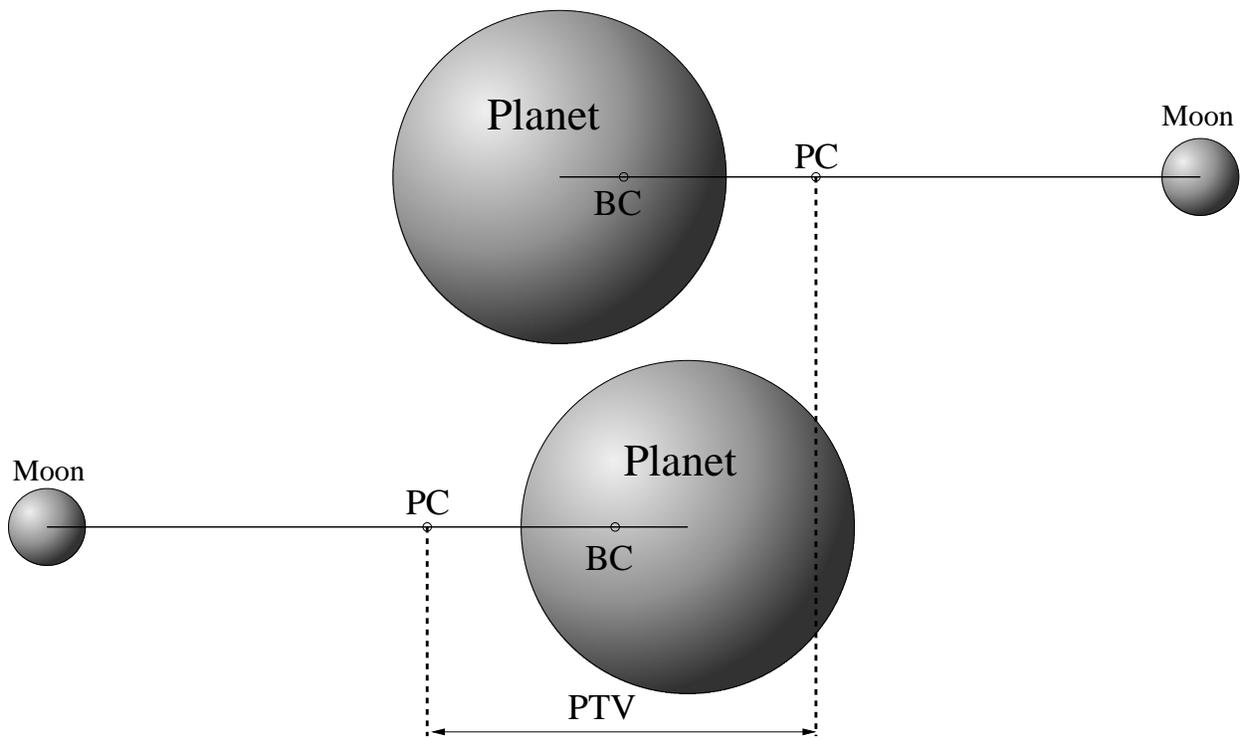}
\caption{The definition of the PTV: the time delay due to the revolution of a theoretical body around the barycenter (Simon et al. 2007). Not to scale.}
\label{ttvp}
\end{figure}

The traditional TTV simply measures the timing variation of the planetary transit and does not consider the tiny photometric effect of the moon. Szab\'o et al. (2006) and Simon et al. (2007) argued against this simplification and proposed a new approach for obtaining the variation of the central time of the transit and calculated the geometric central line of the light curve by integrating the time-weighted occulted flux. They derived a formula, which showed that there is a fixed point on the planet-moon line, which is the so called photocenter (PC in Fig. \ref{ttvp}). In this photocenter an imaginary celestial body cause the same photometric timing effect as the planet and the moon combined together. The motion of this visual body around the planet-moon barycenter leads to the variation of the transit time, which is the photocentric transit timing variation (PTV in Fig. \ref{ttvp} ):

\begin{equation}
PTV \sim \left| \vartheta^2 - \chi \vartheta^3\right|
\end{equation} 
This model takes into account both the photometric and barycentric effect of the moon.

Both methods have advantages and disadvantages. The traditional TTV uses parametric model light curve fitting (e.g. Mandel \& Agol, 2002) to derive the timing variations of the transit. This does not need equally sampled data and the result does not significantly depend on the number of the measurements. Furthermore, it allows characterizing the star-planet pair immediately and gives a physical interpretation of the system. However, the results depend on the model adopted to fit the data. For distorted light curves, blind model fitting can lead to non-physical parameter combinations and ultimately misleading results, hence model independent tests are always very important. Moreover, mapping large parameter spaces for fitting extensive sets of observations can be extremely time-consuming, making a full, detailed analysis nearly impossible. In contrast, the technique of the PTV is a so-called non-parametric method which can provide fast reduction of the light curve shapes even for large sets of data. There is no need to make assumptions on the analytic light curve model, because PTV uses only a numerical summation of time--weighted fluxes in a window with certain width, hence there is no need to increase the number of model parameters for distorted transit curves. It is true for both the TTV and the PTV that their usage is limited by the spatial configuration of the systems. Independently of their size, for close-in moons mutual eclipses occur during the transit, therefore due to the opposite motion of the moon the apparent timing variation will be largely canceled out.

\begin{figure}
\centering
\plotone{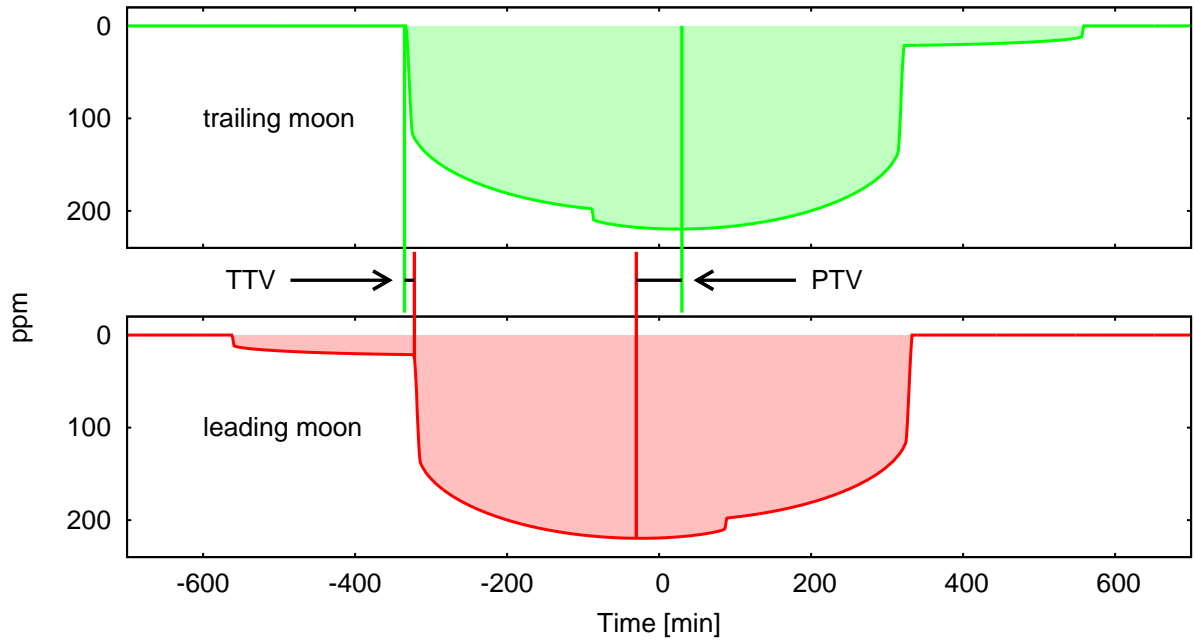}
\caption{The schematic illustration of the differences between PTV and TTV: the PTV effect is larger and appears in the opposite direction than that of the TTV.}
\label{diff}
\end{figure}

Another differences between the techniques is demonstrated in Fig. \ref{diff}, where the green light curve shows a transiting system with a leading moon, while the red one corresponds to the opposite case, when the moon is in a trailing position. As can be seen, the two effects are opposite in sign and the PTV can produce larger signal. Interestingly, one can show from the derived formula of the TTV and the PTV, that the former is more sensitive to the mass of the moon, while the latter gives better constraints on the radius of the moon (Simon et al. 2007). 

To measure the entire PTV effect, the evaluation window should be longer than the transit duration. The size of this window depends on the parameters of the systems (Simon et al. 2012), mainly on the semi-major axis of the moon, therefore on the Hill sphere as well. For our simulation the choice of a window with triple duration width was enough to measure the tiny drops due to the moon for every cases. Even if this small brightness decline on the left or right shoulder of the main light curve cannot be seen directly, they can cause measurable time shift in the transit time (Simon et al. 2007). 

\begin{figure}
\centering
\plotone{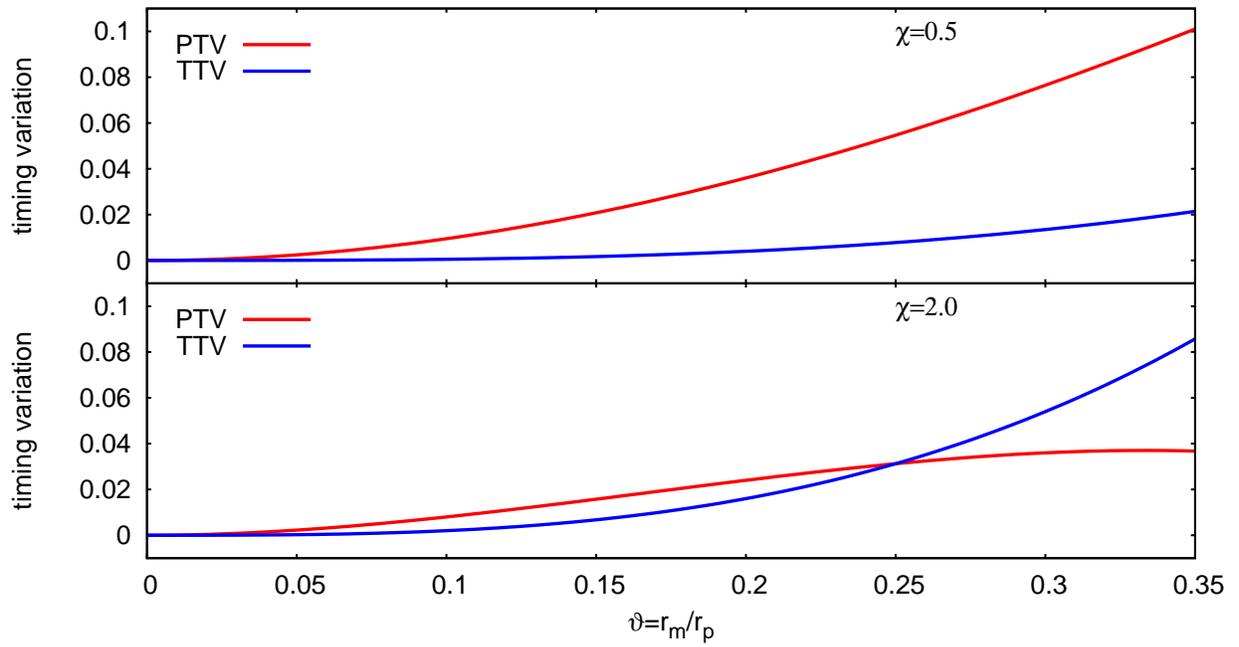}
\caption{Comparison between the half-amplitudes of the timing variations (TTV and PTV). The moon-planet density ratio $\chi$ is indicated in the upper right corners of the panels. The timing variations are expressed in arbitrary units on the y-axis.}
\label{ttvptv}
\end{figure}

We note that the magnitude of these effects depend on the density ratio of the companions. For example, if the density of the moon is twice higher than that of the planet, which can be the case for a gas giant-rocky (or icy) moon system, the TTV dominates only if the size ratio is higher than 0.25. If the moon-planet density ratio is about 0.5, for an Earth-Moon system, the PTV always surpasses the TTV (Fig. \ref{ttvptv}).

Hereafter we will take the advantages of the PTV method to investigate the expected performance of the CHEOPS space telescope. However, to put the results into broader context, we will also compare PTV and TTV in selected cases.

\section{Simulations}

In this paper we used our existing code for precise light curve calculations of transiting exoplanet with an exomoon (for the detailed description see Simon et al. 2009, 2010). Our code takes into account all the dynamical, kinematic and photometric effect, such as non-linear limb darkening, planet--moon dynamical motion, mutual transit, transit timing variation (TTV), photometric timing variation (PTV), transit duration variation (TDV), light curve asymmetry, light curve distortion due to the moon, smearing effect, (in progress: rotating star, starspots) etc. for calculating reliable exomoon light curves and radial velocity curves with the Rossiter-McLaughlin effect.

Although our code is capable of including the full photodynamical model, a comprehensive analysis of discovering exmoons in simulated data is beyond the scope of this paper. The PTV technique kept our investigation as simple as necessary, but yielded fast and robust results for estimating the detectability of exomoons via CHEOPS and define the parameters space of systems, which could be plausible targets for such a detection. 
A search for exomoons in real data, such the HEK Project (Kipping et al. 2014), needs a more thorough analysis where a full photodynamical fit also has to be taken into account in addition to the recent techniques (PTV, TTV, etc.).  

After some recent performance improvements in the calculation speed of our code we developed a decision algorithm with which we can generate and handle large data sets and make detection statistics for such systems. In devising our decision algorithm, we considered the CHEOPS scientific objectives. The main goal of the CHEOPS mission, to be launched by late 2017, is to characterize exoplanets with typical sizes ranging from Neptune down to Earth, orbiting bright stars. The target list will consist of such stars which are already known to host planets, previously detected by accurate Doppler surveys. The planned duration of the mission will be 3.5 years and about 500 target radii will be measured with the precision down to 10\% (Broeg et al., 2013)\footnote{cheops.unibe.ch}.

\subsection{Parameter space}

For the sake of simplicity we assumed circular orbit for the companions and set the impact parameter ($b$) to zero. The semi-major axes were calculated from the third Kepler law.

\begin{figure}
\centering
\plotone{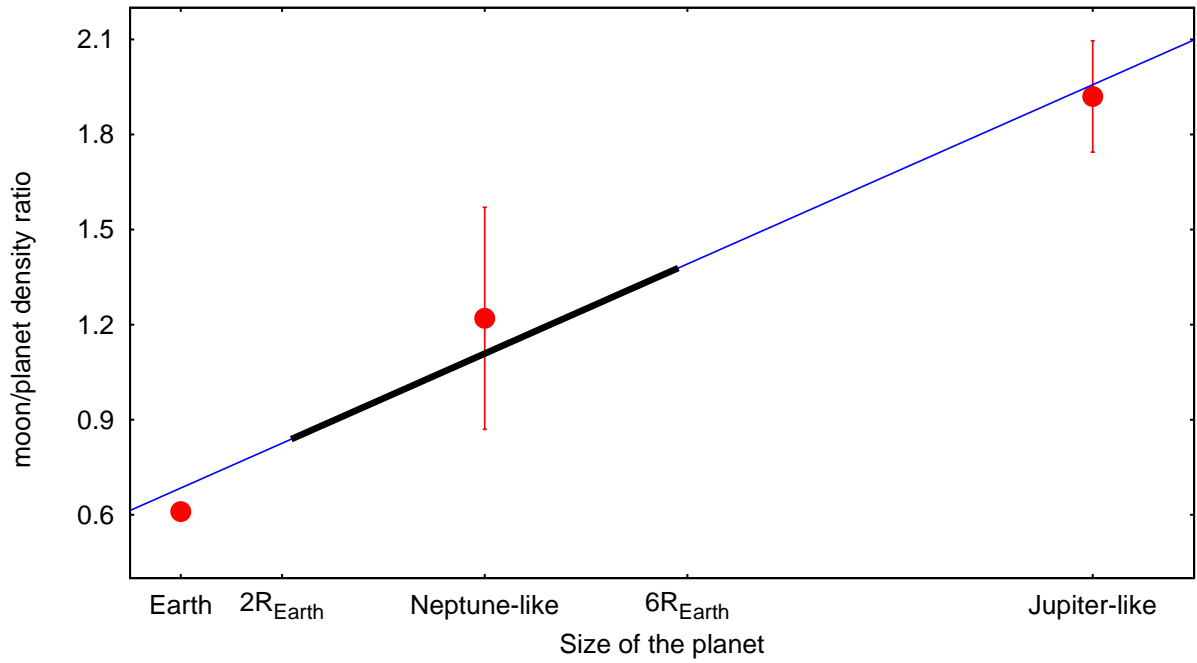}
\caption{The average moon/planet density ratio for the companions in our Solar system (red symbols) and an empirical fit to the data (blue line). The black line refers to the used values in the simulations. The red symbols were calculated for the Earth-Moon pair, for Neptune and its five moons (Titania, Oberon, Umbriel, Ariel, Triton) and for Jupiter and Saturn and their nine largest satellites (Io, Europa, Ganymede, Callisto, Titan, Rhea, Iapetus, Dione, Tethys).}
\label{solar_rd}
\end{figure}

To reduce the large parameter space during our simulations, we provided estimates for some physical variables and defined their intervals. For the sake of simplicity first we chose one solar mass and one solar radius for the host star. We divided our simulations into five main bins with planet sizes ranging from 2-6 Earth radii, with increments of 1 Earth-size. The sizes of the moons were chosen from the interval 0.7 and 1.5 Earth-size, but the planet always had to be larger by a factor of 3 in radius, except for the smallest planet, where the ratio is 2.86 due to the rounding. This restriction in size is necessary to avoid investigating double planets. For PTV calculations we needed the density ratios of the companions. In order to stay within physical reasonable parameter range, we took planetary data in the Solar System (Fig. \ref{solar_rd}). We have considered the Earth-Moon pair, the system of Neptune and its five largest moons and combined the data for Jupiter and Saturn and their nine largest satellites. The result is a correlation between the planet size and moon density, which was used in the simulations: for a fixed planet radius, the density of the moon was calculated using the correlation we see in the Solar System. The actual numbers in the simulations correspond to a moon mass range of 1--50 Earth masses.

Before the detailed simulations we performed some trial investigations for a few representative parameter combinations. These preliminary investigations all agreed in their conclusions that we need at least 15-20 simulated transits to make efficient characterization of a reliable detection statistics for the entire parameter space.

Then we had to set the most important observational constraint, the orbital period of the planet. The upper limit was fixed by the planned operation time of CHEOPS, which is 3.5 years. In order to be able to observe 15-20 transits, the orbital period must be shorter than 70 to 80 days, hence we chose 75 days.

\begin{figure}
\centering
\plotone{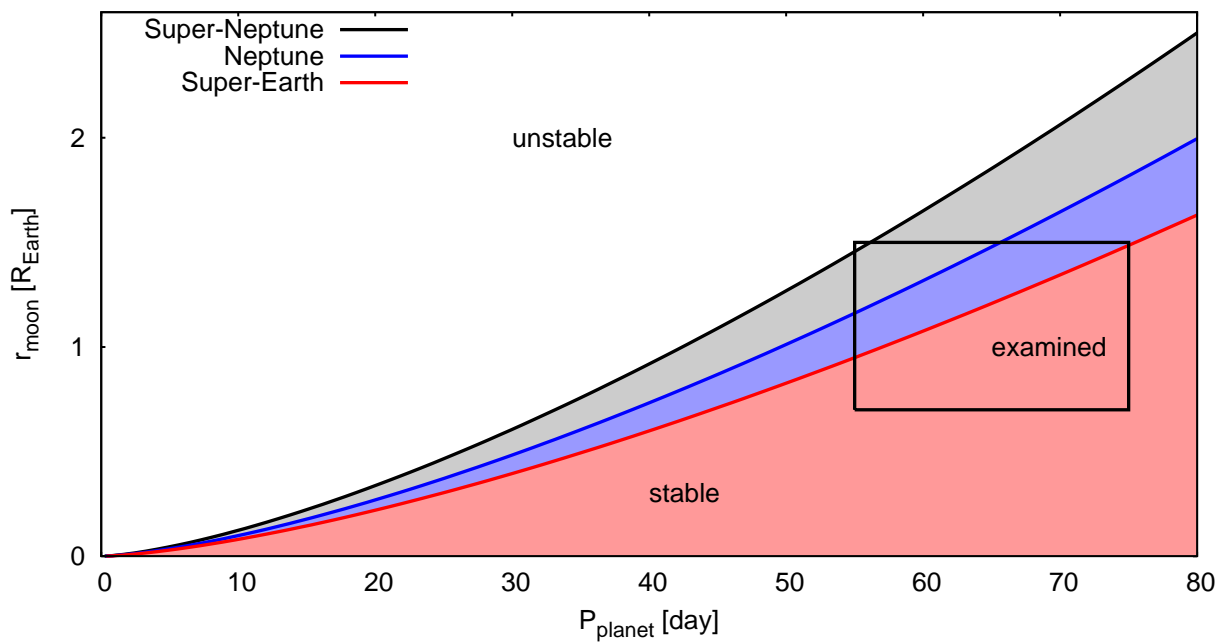}
\caption{Stability diagram for hypothetical moons following the  calculations from Barnes \& O' Brien (2002). The coloured areas correspond to stable moon configurations for 1 billion year in the three indicated cases. The black box refers to the examined range of parameters.}
\label{stab}
\end{figure}

For the lower limit of the orbital period, we took into account the fact that the moons must be both stable for billion years and large enough to produce detectable signal. Barnes \& O' Brien (2002) discussed the tidal stability of satellites and presented a detailed analytic description of the relation between the planet's semi-major axis, the moon's maximum mass and its long-term stability. This can be easily converted into a relationship between the orbital period of the planet and the radius of the moon (Fig. \ref{stab}). For example, if we select a 1.5 Earth-sized moon (the largest one in our simulations) and assume for at least 1 billion years around a Super-Neptune (6 Earth-size), the orbital period of the planet must be longer than 55 days (black curve in Fig. \ref{stab}). Considering this 55 days, the maximum size of a stable moon around a Super-Earth (2 Earth-size) is approximately 0.9 Earth-size (red curve in Fig. \ref{stab}), therefore the choice of 0.7 Earth-sized moon fulfilled this criteria. The rectangular area in Fig. \ref{stab} shows the investigated parameter range.
 
We note that the actual numbers depend on many physical parameters of the system and are only valid for a one solar mass central star and the assumptions of the Barnes \& O' Brien (2002) study, with $0.51$, $10^{-5}$ and $0.36$ set for the tidal Love number, the tidal dissipation factor and the constant fraction, respectively (see also references in Barnes \& O' Brien, 2002). The largest uncertainty in longevity of exomoons lies in the tidal dissipation factor ($Q_p$). While a ($Q_p$ = $10^5$) is commonly used (Weidner and Horne, 2010), Cassidy et al. (2009) suggested values as high as $10^{13}$ for exoplanets. The choice of time for stability has less affect for the calculation, therefore we chose 1 billion years, which is the minimum desirable value for calculating longevity of exomoons (Sasaki \& Barnes 2014), for determining the range of our parameter space. We also note that Sasaki et al. (2012) found that the inclusion of the effect of lunar tides on planet rotation allows significantly longer lifetimes for massive moons. Despite the large uncertainties in the stability calculations, our simulations give a representative view of what we can expect from CHEOPS; for real systems, one must perform as closely matching calculations as possible.

The orbital period of the moon was chosen randomly, where a logarithmic distribution was used. At this point, to exclude the short-time escape of the moon, the system has to fulfill the criterion on the Hill-sphere and the moon's orbit must be beyond the Roche-limit. From the definition of the Hill-sphere one can derive that if we want the moon to orbit closer than the Hill-radius, then the ratio of the orbital periods of the planet and the moon can never drop below the value of $\sqrt{3}$ (Kipping 2009a). The minimum value was 2.5 in our simulations, while the maximum ratio was 20, which fulfilled the criteria on the Roche-limit for every case, too. By choosing the value of 20 for the maximum ratio we can neglect close-in moons, such as Jupiter-Io, but for very large ratios the transit duration and the moon's orbital period will be in the same order of magnitude and mutual eclipses during the transit can occur, hence TTV and PTV will diminish.

The duration of the simulations were equally the total time span of 20 transit (orbits), therefore the shortest and longest simulations ran for 1100 and 1500 days, respectively. The cadence rate was set to 60 seconds. The total number of differently configured systems was 260,000. This number came from 5 main planet-moon combinations (see Table \ref{syspar}), where parameters were randomly chosen from the next parameter space: the size of the moon and the ratio of the orbital periods. In these main bins the number simulations were proportional to the width of the interval of the moon size, where every 0.1 step in the moon size means 10,000 runs. To make the detection statistics for each configuration, we performed another 3000 simulated observations by using the bootstrapping technique. The only difference among these runs was in the time sequence of the noise and half of these simulations were performed without moons, which meant the null events, i.e the detection limit for our decision algorithm (see later). 

\subsection{Noise model}

Another important component of the simulations is the applied noise model. The noise was generated by the {\it CHEOPS Noise Simulator}, which was still in development phase during this study, therefore a preliminary noise model was used. We note that as the design of the mission progresses, noise estimates will become available. Because the detailed model description and documentation of {\it the CHEOPS Simulator} is not available for public use, here we can give a short description about its current state and the considered noise budget.

\begin{figure}
\centering
\plotone{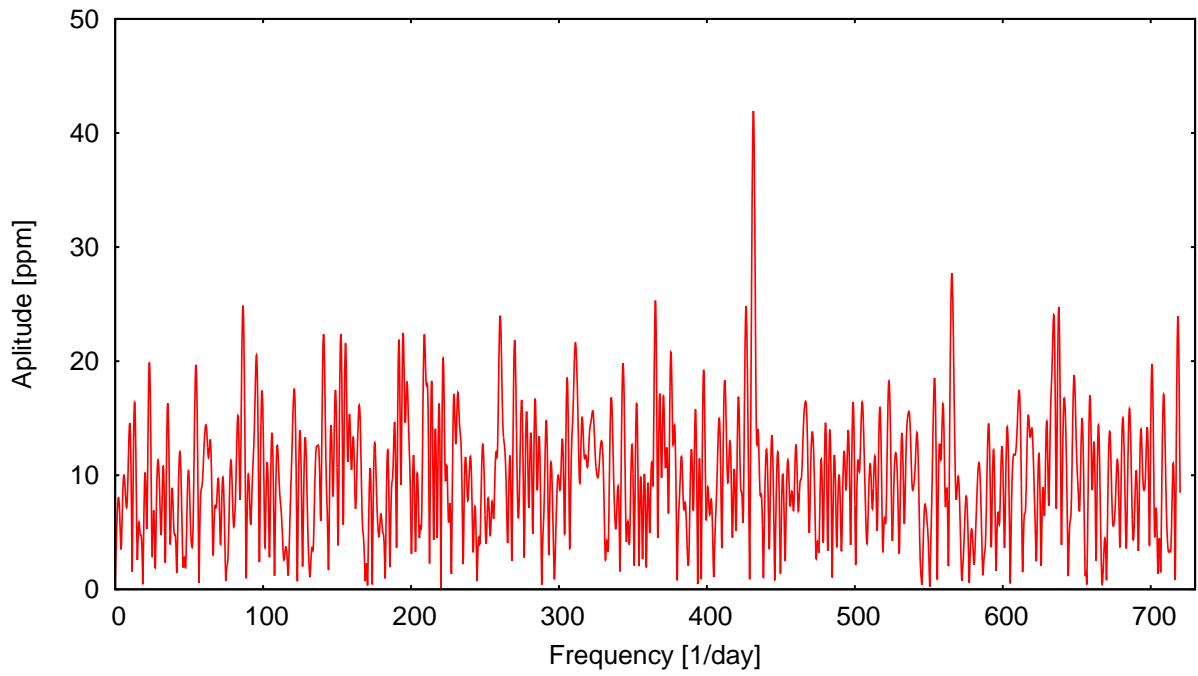}
\caption{The Fourier spectrum of the applied non-Gaussian noise. There can be seen that there is a periodic component around the frequency of 430/day.}
\label{noise}
\end{figure}

First {\it the CHEOPS Simulator} produces images by calculating the expected PSF of a star. At this point the spectral type of the star, the temperature of the CCD and the telescope throughput are considered. The measurement points are obtained by performing a simple aperture photometry. At the time of the simulations the {\it CHEOPS Noise Simulator} could consider the following noise sources: shot noise; read-out noise; quantization noise; sky background (Zodiacal light); temperature induced variability of the dark current; Quantum Efficiency variation of the detector with temperature; gain variation of the proximity electronics; flat-field errors; bad pixels; stray light; cosmic ray effects, jitter due to the pointing accuracy. The stellar noise was simplified by adding a value drawn randomly from a Gaussian distribution. The first four were treated as white noise. The other sources are related to systematic or environmental noises, which are non-random and likely to be predominantly periodic possibly synchronized with the period of the chosen orbit. The Fourier spectrum of the applied noise can be seen in the Fig. \ref{noise}, where a peak around 430 $1/day$ shows a periodic component indicating that our simulated noise differ somewhat from the Gaussian distribution. We note that the predominant source per exposure is the shot noise, however, its time average decreases rapidly and the environmental and systemic noises can have more significant effects.

Lewis (2011,2013) studied more thoroughly the effect of a realistic stellar noise for the detectability of exomoons via PTV method. It was found that in the case of a filtered solar noise the size of a detectable moon increases by a factor of 1.5 (Lewis, 2011). Our analysis showed that the inclusion of the systematic and environmental noises also results in a growth in the moon size with 10-20\% compared to the white Gaussian noise. Despite this sensitivity to noises of the PTV method, this technique can still produce a signal, which can usually be larger than those of the other methods (TTV, TDV, etc.) and can offer a statistically fast and still robust evaluation. Therefore using PTV is still a reasonable choice for analyzing huge simulated data sets, selecting targets and estimating the possible detection rate of exomoons via CHEOPS.

Because the development of the {\it CHEOPS Simulator} is still in progress, the implementation of the more realistic stellar noise is also under development. We had to adjust the preliminary noise by multiplying with a certain factor to meet the requirements of the photometric precision, which is 20 ppm during a 6 hours of integration time (CHEOPS Definition Study Report\footnote{http://sci.esa.int/cosmic-vision/53541-cheops-definition-study-report-red-book/}, 2013). Given that the planned magnitudes of the CHEOPS targets range from 6 to 12, we fixed the apparent brightness of the simulated signal at 9$^{th}$ magnitude, with the corresponding noise characteristics.

It is known that there are several other unknown factors that can contribute to the noise budget, such the celestial position, the brightness, the real stellar noise, etc. of a target, therefore the present approximation of the adjusted noise is the main limitation factor of our investigations. Because the primary aim of this investigation is not to discover and confirm exomoons in the real data, such as the HEK Project (Kipping, 2014), but to present a feasibility study for the CHEOPS space telescope, which is still under development, and to propose possible targets with certain detection rate and to answer whether this instrument are capable of an exomoon detection, the more sophisticated modeling of the total noise budget is not a goal of this paper.

\section{The decision algorithm and the procedure}

\begin{figure}
\centering
\plotone{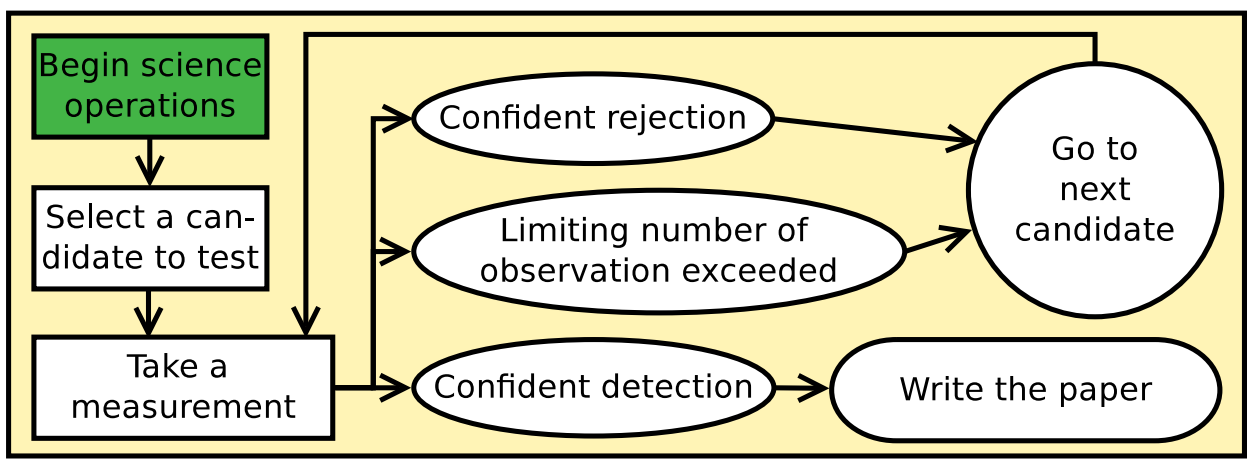}
\caption{The concept of decision algorithm. The observation cost must be minimized by rejecting noisy measurements and finding the minimum number of transits to observe for confident detection.}
\label{dec}
\end{figure}

The discovery of an exomoon requires multiple transit measurements to reduce the noise level without wasting time to systems that have no signal or are clearly detected. Consequently, the search must be optimized. We integrated our exomoon simulation code (Simon et al. 2009, 2010) into a newly developed decision algorithm that is capable of searching the most efficient detection index from simulated data, along a pre-set false alarm levels and desired lower limits. The concept of this algorithm was speeding up the process of the decision whether the tested system is a potential candidate or the detected signal cannot be distinguished from the statistical noise. In addition this has to be done for a limited number of transit observations (Fig. \ref{dec}). 

We used the bootstrapping technique to determine the threshold levels for detection/rejection based on the measured PTV values. For this we performed 3000 simulated observations for every given system configurations as follows. First, we computed the PTV for the null event from 1000 runs, where only planetary transits without moons were simulated, consequently the ``measured'' PTV comes purely from the statistical noise. After sorting these 1000 PTV values, the algorithm selects the top $n^{th}$ point as the detection threshold according to the desired a posteriori output. Second, we performed another 1000 bootstrapped simulations with the same planet + moon configurations to determine the minimum PTV amplitude that can be produced by and expected from the given configuration. After sorting the measured PTV values the algorithm selects here the bottom $m^{th}$ value as the rejection threshold. Finally, we made another 500 runs without moons and 500 runs with moons randomly to calculate the detection statistics for the actual configuration. With the capability of adjusting threshold levels we can balance between the "speed" of the decision and the number of false alarms. The lower detection threshold means less observing time and higher detection rate, but larger false alarms. On the other hard, pushing the rejection threshold higher makes  the chances of loosing a true system higher, therefore both the detection rate and the number of false positives is lower. We note that despite the adjustable threshold levels (i.e. the freedom of choosing different $n$ and $m$ values), we have chosen fix values, namely $n=980$ and $m=20$, to speed up the algorithm, and decrease the computation time for 260,000 systems. These choices translate to a false-alarm probability of 2\%, which corresponds to about 2.5$\sigma$ (one sided) if we assume a normal detection statistics. The given significance level is an important input parameter in determining the detection strategy, and can be set to an arbitrary level (e.g. 4$\sigma$) if the problem requires so.

\begin{figure}
\centering
\plotone{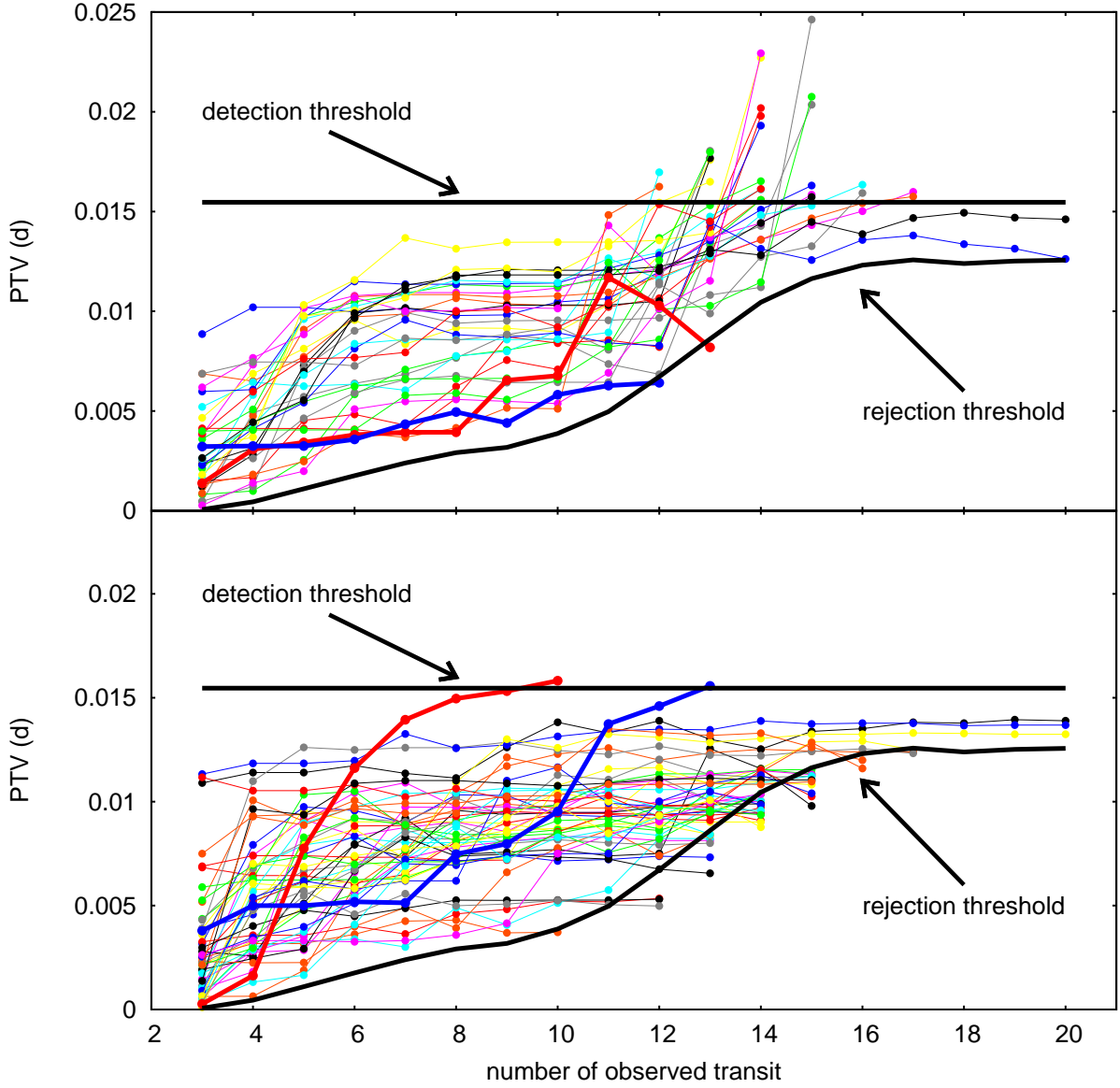}
\caption{Two examples showing thirty evolution tracks from the detection statistics of a simulated system. These coloured tracks show the maxima of the PTVs as they are evolving with the number of observed transits, while the black ones mean the upper and lower boundaries of the decision making, where the given systems will be marked as detection or rejection. The upper/lower panel illustrates as PTV values of the systems with/without moons are crossing the detection/rejection thresholds. Both panels also have false detection/rejections indicated by thicker colour lines.}
\label{evotrck}
\end{figure}

During the evaluation process for a system, the algorithm computes the central time for each transit light curve and fits a linear model to the set of observed minus expected central times. The residual of the linear fit define the PTV values. Then it determines the rejection and the detection thresholds with the bootstrapping technique, and subsequently evaluates the simulated observations. The rejection threshold, which is the expected minimum amplitude of the PTV for the given system, is calculated as the maximum deviation from the linear fit. As expected, this value increases with the number of observed transit as more and more points are measured and the period of the PTV becomes apparent. While determining the detection threshold, we expect null signal for the amplitude of the PTV, therefore zero variation for the differences between the expected and observed central times. (A planet without any external influences by other companions, such as other planets and moons can be considered strictly periodic within the studied time-scale of 3.5 years.) To determine the maximum signal due to the statistical noise, we searched for only the maximum deviation from zero. Since the noise level was constant during the analysis, the maximum differences between the expected and observed central times, thus the PTV, are independent of the number of observed transits, therefore the detection threshold remains constant (Fig. \ref{evotrck}).

By increasing the number of measured transits the algorithm  re-evaluates the PTV and compares that to the thresholds (Fig. \ref{evotrck}). If a systems exceeds the detection limit or drops below the rejection limit, it will be either classified as a potential candidate of hosting moon or rejected after a certain number of observed transits. If the target would need more observation than a limited number to be classified, it remained in the undecided state. Two examples are shown in the Fig. \ref{evotrck}. The upper panel refers to systems with moons, where the evolution of the PTV amplitudes show ascending trends, while the systems without moons evolve in horizontal direction except two false detections indicated by thicker lines.

The final step of the algorithm arrives when a certain system gets classified into one of the three categories: detection, rejection, undecided. To optimize the observing strategy, we need to know how much of the limited instrument time should be spent in order to get an unambiguous classification. For this, we took the average classification time for the 500 simulations with and 500 simulation without a moon. This approach is analogous to the case when the probability of the existence of a moon is 50\% and we expect a clear yes/no/indeterminable answer from the algorithm (see  T$_{50}$ in Table \ref{syspar}). 

Here we note for the sake of interest: in the first part of our simulations we analyzed general cases by splitting the size of planets into five bins. For each case the number of the simulations was proportional to the width of the interval of the moon radius, therefore the total number of simulated configurations was 260,000 (Table \ref{syspar}). Each run consisted of the 3000 simulated observations (see above) and every simulation had 20 transits. This means in total about 15.6 billion calculation of the PTVs.

\section{CHEOPS performance and detection statistics}

Our investigation of the potential CHEOPS performance in detecting exomoons consisted of two parts.  The first part was done in a general way, in which the systems were divided into five main bins by defining the size of the planets and randomizing the other physical parameters. The second part consisted of candidates which already have radial velocity measurements (data taken from The Extrasolar Planets Encyclopaedia\footnote{explanet.eu}) by fulfilling our simulation criteria and the pre-defined restrictions for the physical properties of the companions. To make assumption to the boundaries of expected radius of these planets we used observational and theoretical results form Weiss \& Marcy (2014) and Swift et al. (2012), who studied the correlation between the masses and the radii of exoplanets. Then we simulated transit observations by taking values randomly for the moon sizes and for the ratio of the orbital periods.

The probability of detecting exomoons around both the simulated systems and the selected candidates was obtained by averaging the detection statistics of individual systems. This a posteriori output highly depends on the pre-set levels of the threshold during the bootstrapping analysis. Forcing fast decisions can yield to a large number of false hits, while choosing high confidence levels for detection and rejection can leave a lot of systems in undecided states thereby wasting observing time.

\subsection{Detection statistics for simulated observations}

We summarized the physical parameters (first six columns) of the simulations and their detection statistics (last four columns) in Table \ref{syspar}. The number of discoverable exomoons (column 7) increases with the size of the planet and almost reaches 90\% in the case of 4 Earth-sized planet at less than 2\% false alarm rate (column 8). This result is a bit surprisingly because we expect smaller PTV values in the cases of larger planetary sizes according to equation (2) and therefore fewer detection. As our probability tests showed the simple explanation is that the signal to noise ratio (in this case the ratio of the transit depth and the magnitude of the noise) is increasing with the radius of the planet and this effect is stronger than the decrease of the PTV, so we get more hits even if the value of the PTV is smaller. For a planet with 4 Earth-size, the average observing time (column 9), in order to detect the signal of an existing moon, decreases with the planetary size and is about 6.7 days. (We note that for most of the systems in the explored parameter range, the average transit duration is about 8 hours, hence one full transit observation requires 1 day of monitoring. Therefore, there is a close correlation between the required number of observed transits and the total number of days spent on a target.)

\begin{figure}
\centering
\plotone{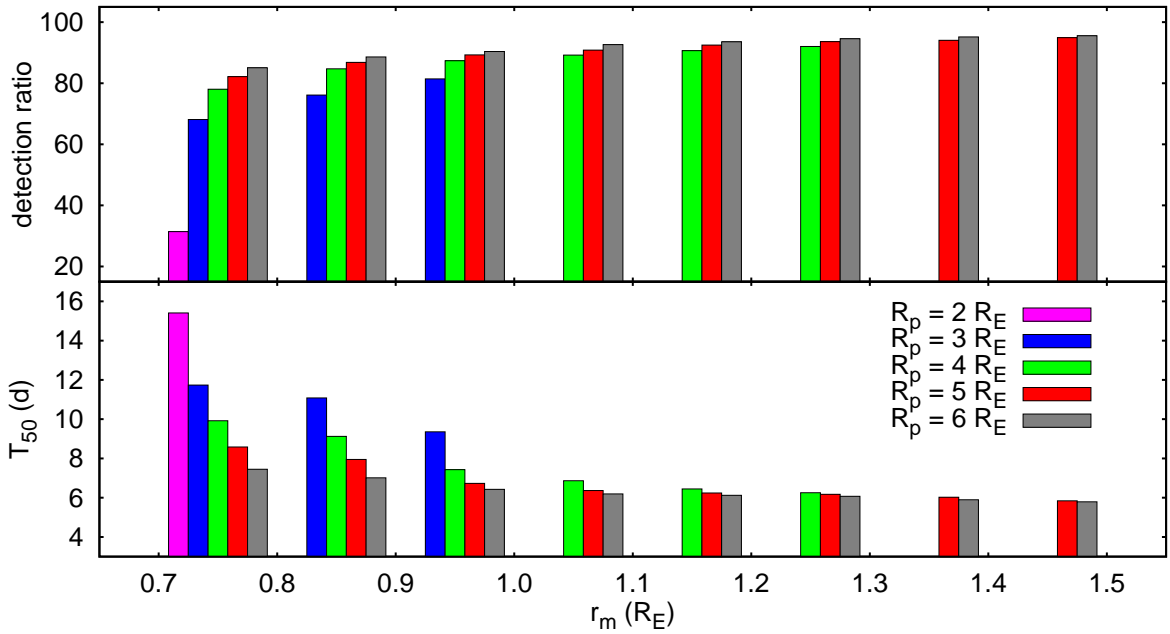}
\caption{The detection statistics of the simulated transiting planets with different-sized moons ($r_m$). The upper panel shows the ratio of the detected systems, which can be interpreted as the detection probability if we assume that these moons can exist around these planets. In the lower panel one can see the average observing time which must be spent until the measured systems will be classified in one of the three categories.}
\label{moon_stat}
\end{figure}

Fig. \ref{moon_stat} shows how the detection statistics is changing with the size of the moon and the planet. It is surprising that we have more than 70\% chance for a detection of an 0.8 Earth-sized exomoon if the host planet is bigger than 3 Earth-size. The observing time starts dropping rapidly below Earth-sized moons but does not exceed 10 days (less than 1\% of the expected operation time of CHEOPS) for planets that are bigger than Neptune. 

\begin{figure}
\centering
\plotone{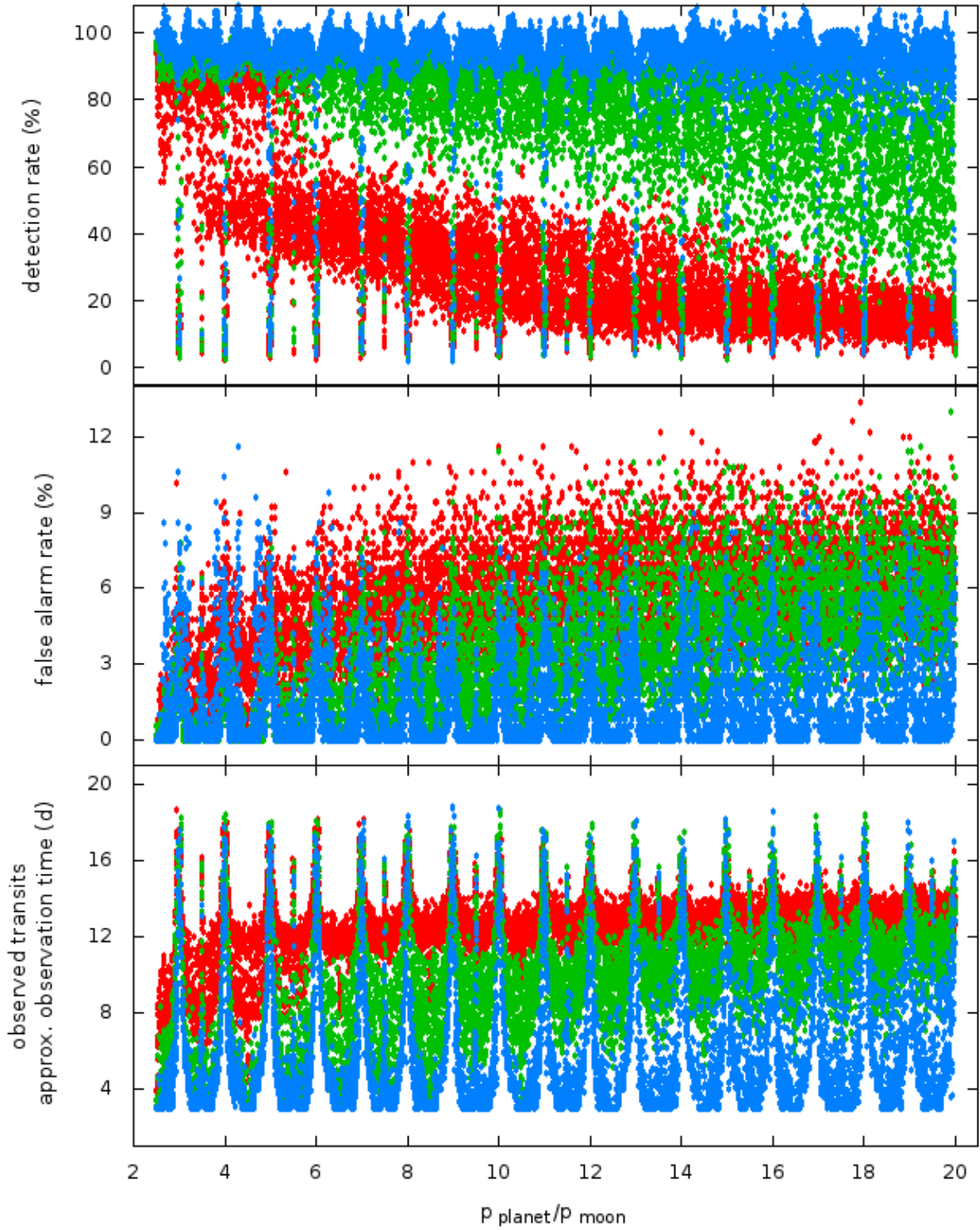}
\caption{Detection statistics for three representative planet sizes. The red, green and blue colours correspond to 2, 3 and 6 Earth-radii, respectively. The upper panel shows the ratio of the detected systems, the middle panel corresponds to the false hits, while the lower panel gives us information about the minimum number of the necessary transit observations for a successful detection, which equals approximately the total measurement time in our cases.}
\label{detstat}
\end{figure}

Another aspect of the detection statistics is illustrated in Fig. \ref{detstat}, where the variations are plotted against the ratio of the orbital periods for three representative planetary sizes. The figure clearly shows that success rate increases with the size of the planet (indicated by colours), while the required observing time decreases. The large spread within the same colours is caused by the fact the simulated moons have different sizes while the planet sizes were fixed. Also there is a clear repetitive pattern for the orbital resonances, when the periods are commensurable. In such cases there is a drop in the detection rate, which is caused by those configurations in which the planet and the moon transit the stellar disk almost always in the same relative phase, so that there is no PTV effect. For such a reason 100\% cannot be reached for the average detection rate (``adr'' in Table \ref{syspar}). Another finding is that there is a decreasing trend for larger period ratios, that is for the short-period close-in moons. This agrees well with the result of Simon et al. (2012).

We have to note that the detection rate cannot be interpreted alone without considering the false alarm rate and the time needed for the detection. Neither the false alarm rate nor the observing time are constant for different configurations. For example, the variation of the false alarm rate can be seen in the middle panel of Fig. \ref{detstat}, where the values change from zero to twelve percent. In the case of commensurable orbital periods it is much higher than elsewhere. When such a configuration with an orbital resonance is being examined, the rejection level will be much lower because the expected minimum value for the PTV is zero, hence it does not differ from the statistical noise. This means that many false signals can survive the first steps, which increases the chances of their detections in the later steps. This can be noticed as peaks at certain ratio of the orbital periods in the middle and lower panels of Fig. \ref{detstat}. The level of the rejection threshold can be pushed upward by choosing higher $m$ value or feed-backing the false alarms to the algorithm, which can automatically raises the rejection threshold and making a new evaluation as many times as the false alarm reaches the desired levels. Our code is capable of such training but for the sake of saving computation time we simplified our investigations by setting the thresholds to a fix $n^{th}$ and $m^{th}$ value according to the result of the bootstrapping. 

\begin{figure}
\centering
\plotone{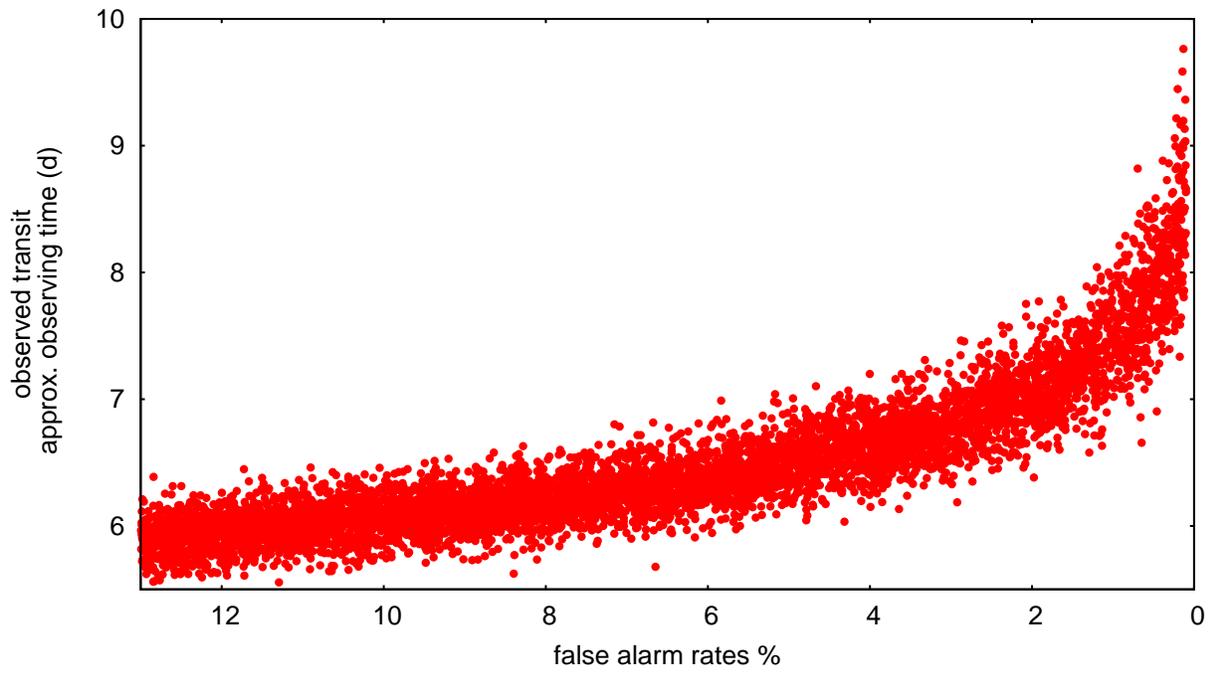}
\caption{The relationship between the false alarm rate and  the number of observed transit for an example system (Neptune-sized planet, Earth-sized moon).}
\label{fap_dsp}
\end{figure}

To illustrate the relationship between the false alarm rate and the number of observed transit, we made a separate simulation, where the parameters of the systems were the same, only the level of the false alarm rate was varied. The results showed that pushing the false alarm rate for the detections towards the lower values increases the observing time nearly linearly until 2\% then below this, the observing time increases nearly exponentially (Fig. \ref{fap_dsp}). Repeating this simulations for other configurations the relation was found to be similar, only the number of transits to be observed varied.

\subsection{Known exoplanets with radial velocity measurements}

We repeated our simulations by using real physical parameters from The Extrasolar Planets Encyclopaedia, where 428 planetary systems are listed, which were discovered by radial velocity measurements. Because the results of the simulated observation showed that 6-10 transit measurements is enough for reaching decision in most of the cases, we selected planets that have orbital period between 55 and 100 days, so we will not exceed the planned observing time, neither in the case of the longest orbital period. We chose those planets which had masses larger than 5 but did not exceed the 70 Earth-masses. After applying these restrictions, 10 systems remained, mostly similar apparent brightness range than what was assumed previously. For estimating the expected sizes of the planets we used the result of Weiss \& Marcy (2014), who studied mass-radius correlation of 65 exoplanets smaller than 4 Earth-size and with orbital periods shorter than 100 days and that of Swift et al. (2012), who derived relations for different classes of material in wider range of mass and radius. The other parameters of the moon have been chosen randomly as described previously. The properties of the systems and the results for their detection statistics are listed in Table \ref{canpar}. 

Despite the small number of the examined exoplanets and the shorter observing time (not more than 10 light curves) we got more than 80\% successful detection rate with less than 1.5\% false alarm probability and with no more than 7 days of observing time for half of the cases (5,6,8,9,10 in Table \ref{canpar}). This means that if we do not stick to this strict restriction (allowing larger moons, more observing time) the success rate could be very promising, so that CHEOPS could have good chances for an exomoon detection. Nevertheless, we have to note here that we assumed that we can detect transit, but obviously it is not possible in every case, because the planets to be observed were discovered by Doppler-method and the spatial configuration of the systems can be different from the ideal case. 

We can notice that two planets (5,6) are members of multiple systems, therefore they can show TTVs, which can bias the signal of the PTV. A possible solution for getting unbiased PTV values is to make a time correction by shifting the measurement points of the transit with the appropriate values of the TTV. If the signal of the PTV is still present than there should be another signal in the light curve which is due to an exomoon or sign similar to an exomoon. To decide the origin of the detected PTV signal we need to consider additional detection techniques, which has to show the same result as the PTV did to confirm a presence of probable exomoon. In the following we present two semi-independent methods.

\section{Possible confirmation methods}
\label{conf_methods}

Although there is no consensus about the ``best'' method for an exomoon discovery, the classical approach can be used, where independent techniques provide independent confirmations. Here we propose two simple semi-independent methods for confirming if the detected PTV signals are due to an exomoon. We use the term `semi-independent' because all of these techniques quantify different indirect signals from an exomoon but use the same photometric measurements and are based on the analysis of the same transit light curves. We note that for a reliable confirmation of an exomoon in the real data we have to consider all the possible detection technique including also the photodynamical fit.

\begin{figure}
\centering
\plotone{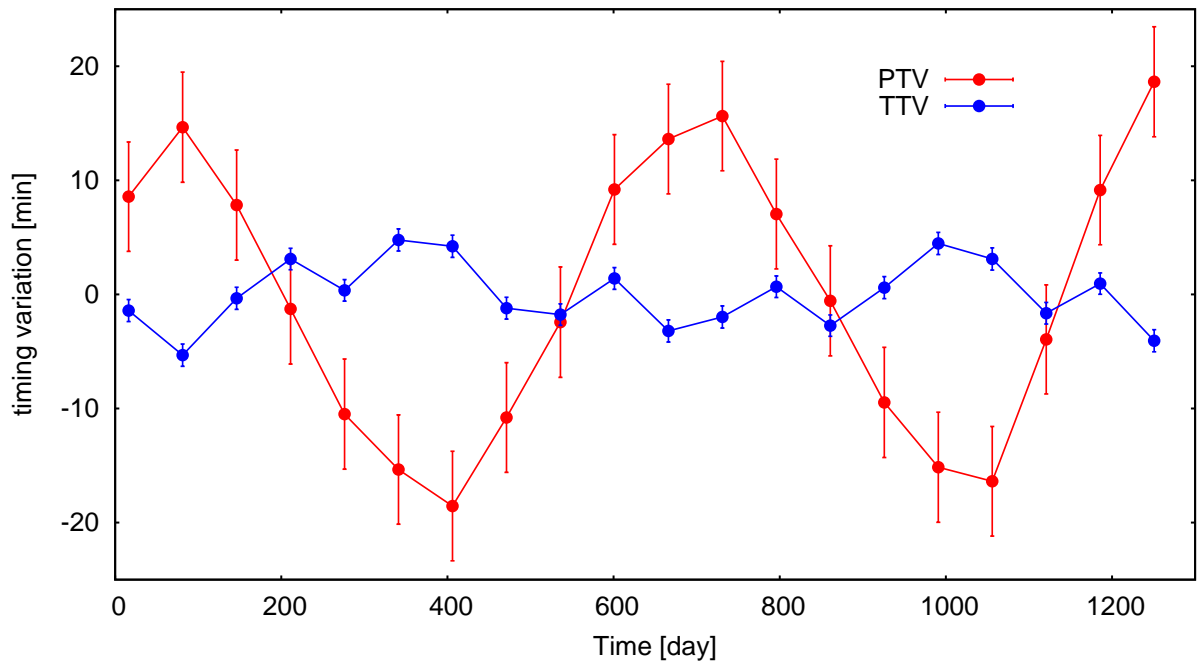}
\caption{The PTV and the traditional TTV are anti-correlated when the distortion of the transit curve is due to the exomoon. For all other kind of modulations, the sign and the magnitude of the variations are comparable. This means that if the data quality is enough for calculating both variations and their observed directions are opposite in sign then that can confirm the presence of an exomoon.}
\label{ttv_ptv_oc}
\end{figure}

As we already discussed the differences between the traditional TTV and the PTV in Sect. 2, here we only emphasize again that the amplitudes of the produced signals differ from each other. A previous analysis showed that the PTV can produce stronger signal in opposite phase to the TTV. To illustrate this we simulated a transiting system with half Jupiter-sized planet and 1.2 Earth-sized moon around a star with a solar mass and radius. The orbital periods of the planet and the moon were 64.973 and 30.862 days, respectively. Both timing variations can be seen in Fig. \ref{ttv_ptv_oc}, which shows that the PTV's signal has much larger amplitude and it runs oppositely in sign. This latter one is because the photocenter is on the opposite side of the barycenter than the center of the planet (compare Figs. \ref{ttvb} and \ref{ttvp}). The PTV measures the timing variation of the photocenter which can be derived by integrating the brightness decline both from the moon and the planet. Although the planet can produce deeper light curve, the moon is located far from the planet and transits much later (earlier), so the longer time basis can amplify its small photometric effect. In contrast, the TTV measures only the shift of the main light curve, which is linked to the distance between the barycenter and the center of the planet. If both variations and their opposite behavior can be measured then it can provide a good evidence for the existence of an exomoon. We note that because the effects of the moon in both sides of the light curve need to be measured, the evaluation window of the PTV technique is much larger than just the main transit of the planet. In practice the light curve should be evaluated in a window that is at least three times wider than the transit duration. 

\begin{figure}
\centering
\plotone{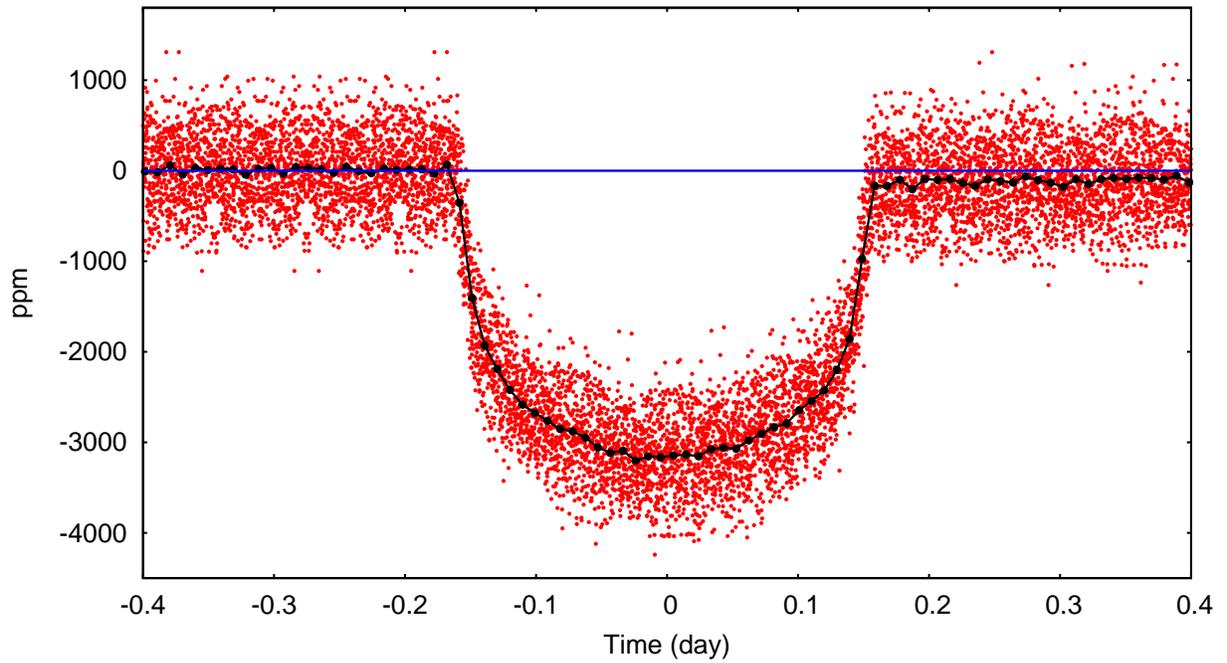}
\caption{Mirroring individual light curves according to their PTV values, then folding (red dots) and averaging (black line and points) them can reveal the tiny brightness drop from the moon that is located at large elongations for most of the time. The right wings of the light curve is significantly deeper than the left side, which can be clearly seen by comparing the level of the black points to the zero (blue) line.}
\label{ptv_fold}
\end{figure}

The other method is similar to Heller's (2014) idea who presented a sensitive method to detect exomoons. This approach is based on the fact that moons appear more often at larger separation from their planet. In that case a photometric orbital sampling effect appears in the phase-folded light curve causing a small decline in stellar brightness immediately before and after the planetary transit. In our technique the PTV actually measures the delay in the central time of the transit compared to the expected value, which is defined by the orbital period of the planet. Depending on the sign of the PTV, the tiny distortion from the moon appears on the opposite side of the light curves. After mirroring the appropriate light curves and folding them the tiny effects of the moon gather on one side of the light curve. After applying a simple moving average to the phase-folded data one can reveal and confirm whether there is a tiny brightness drop caused by the exomoon (Fig. \ref{ptv_fold}).

\section{Summary and conclusions}

In this paper we investigated the performance of the CHEOPS space telescope in searching and detecting exomoons. In all of our simulations we used the technique of photocentric timing variation (PTV), which was introduced by Szab\'o et al. (2006) and Simon et al. (2007). By comparing to the traditional TTV we concluded that both methods measure the variation of the central time of the transit, but the PTV is more sensitive to the tiny light curve distortion due to a moon and produce opposite signal than the TTV. This anti-correlated behavior occurs only in that case when an exomoon is present, therefore can help the confirmation process by itself and as well as at the mirrored folding technique.

We developed a decision algorithm and integrated it to our existing code (Simon et al., 2009), with which we generate billions of light curves and calculated the same amount of PTV values to determine the number and the ratio of systems with detectable exomoons. The idea was to develop a code that can make a fast decision about any observed system, can recognize potential moon-host candidates and reject those which exhibit only such a signal that cannot be distinguished from statistical noise. 

After applying some restrictions to the physical parameters of the planets and moons by fulfilling the criteria of the CHEOPS mission, we divided the possible configurations into five general bins on one hand, and, on other hand, we selected appropriate candidates from the database of {\it The Extrasolar Planets Encyclopaedia}. Taking into account stable moon configurations we randomized the moon parameters in pre-defined intervals and monitored how the maximum of the PTV was evolving with the observing time, how many observed transits are necessary to detect or reject a system and what are their ratio. Our results showed that above Neptune-sized planets and Earth-sized moons, the expected ratio of the detected systems exceeds 85\% with less than 2\% false alarm rate and less than 8 days observing time. Among the 10 analyzed candidates, which were discovered by Doppler-method, we found five, for which the presence of a moon would be detected 80\% probability with less than 1.5\% false alarms and less than 6.5 days (six transit observations) measurement time. 

In the last years there were several attempts to identify signals from exomoons in the Kepler data (Szab\'o et al. 2013; Kipping et al. 2012, 2014, HEK Project) but there is still no firm evidence of such a detection so far. The next space telescope that is designed to observe exoplanetary transits will be CHEOPS. Although its light collecting capability will be much smaller than that of Kepler, the dedicated observing strategy and brighter targets rate will make it at least equivalent or better photometric instrument Kepler was. Our results indicate that there is a good reason for optimism in regards of detecting exomoons using CHEOPS.

\acknowledgments

\section*{Acknowledgements}

This project has been supported by the Hungarian OTKA Grants K83790, K104607, the ESA PECS Contract No. 4000110889/14/NL/NDe, the "Lend\"ulet-2009 Young Researchers" Program of the Hungarian Academy of Sciences and by the City of Szombathely under agreement No. S-11-1027. The funding of AES from the Sciex Programe of the Rector's Conference of the Swiss Universities is gratefully acknowledged.

\clearpage

\clearpage

\begin{deluxetable}{lcccccccccc}
\tabletypesize{\scriptsize}
\tablecaption{Parameters of the companions around a star with one solar mass and one solar radius).\label{syspar}}    
\tablewidth{0pt}         
\tablehead{
\colhead{name} & \colhead{N} & \colhead{$R_{p}\; (R_E)$} & \colhead{$r_{m}\; (R_E)$} & \colhead{$\rho_{m}/\rho_{p}$} & \colhead{$P_p (d)$} & \colhead{$a_p (AU)$} & \colhead{adr (\%)} & \colhead{afar (\%)} & \colhead{T (d)} & \colhead{T$_{50}$ (d)}
}
\startdata
sim2 & 10,000	&   2.0 & 0.7-0.7	& 0.80	& 55-75 & 0.284-0.349 &  31.41 & 5.42 & 11.72 & 15.41 \\    
sim3 & 30,000	&   3.0 & 0.7-1.0	& 0.95	& 55-75 & 0.284-0.349 &  75.16 & 3.42 & 8.98  & 10.52 \\
sim4 & 60,000	&   4.0 & 0.7-1.3	& 1.10	& 55-75 & 0.284-0.349 &  87.28 & 1.89 & 6.74  & 7.43  \\
sim5 & 80,000	&   5.0 & 0.7-1.5	& 1.25	& 55-75 & 0.284-0.349 &  90.54 & 1.52 & 5.92  & 6.61  \\
sim6 & 80,000	&   6.0 & 0.7-1.5	& 1.40	& 55-75 & 0.284-0.349 &  91.97 & 1.39 & 5.55  & 6.31  \\
\enddata
\tablecomments{The ratio of the orbital periods was varied between 2.5-20. The moon was always smaller than one third of the planetary radius (except the case of 2.0 Earth-sized planet) and the moon always fulfilled the Hill- and the Roche-criteria. Symbols: N: number of the simulations; $R_p$:radius of the planet; $r_m$: radius of the moon; $R_E$: Earth-radius; $\rho_{m}/\rho_{p}$: density ratio of the moon and the planet, calculated from the Solar System correlation depicted in Fig \ref{solar_rd}; $P_p (d)$: orbital period of the planet; $a_p (AU)$: semi-major axis; adr (\%): average detection rate; afar (\%): average false alarm rate; T: average observing time spent on the detected systems in days; T$_{50}$: average observing time spent on measuring a system until it was classified into one of the three categories: rejected, detected or undecided cases; the maximum allowed number of the observable transits was set to 20.}                              
\end{deluxetable}

\begin{deluxetable}{rlcccccccccc}
\tabletypesize{\scriptsize}
\tablecaption{Parameters of the selected CHEOPS candidates.\label{canpar}}    
\tablewidth{0pt}         
\tablehead{
\colhead{} & \colhead{$M_p (M_E)$} & \colhead{$R_p^* (R_E)$} &\colhead{$r_{m}\; (R_E)$} &\colhead{$\rho_{m}/\rho_{p}$} &\colhead{$P_p (d)$} &\colhead{$a_p (AU)$}  &\colhead{adr (\%)} &\colhead{afar (\%)} &\colhead{T (d)} &\colhead{T$_{50}$} &\colhead{(d)}
}
\startdata
1. & HD 20794 d  &       4.8   &            2.0   & 0.7 &          0.80 &     90.3  &       0.350  &   44.95 & 2.56 & 8.25  & 9.56    \\
2. & GJ 667C d   &       5.1   &            2.0   & 0.7 &          0.80 &     91.6  &       0.276  &  45.02 & 2.70 & 10.42 & 12.34    \\
3. & GJ 682 c    &       8.7   &        2.0-3.0   & 0.7-1.0 & 0.80-0.95 &     57.3  &       0.176  &  52.70 & 2.26 & 9.84  & 11.23    \\
4. & HD 134606 c &       12.1  &        3.0-4.0   & 0.7-1.3 & 0.95-1.10 &     59.5  &       0.296  &  64.22 & 2.29 & 5.79  & 6.62        \\
5. & HD 20781 c  &       15.8  &        3.0-4.0   & 0.7-1.3 & 0.95-1.10 &     85.1  &       0.346  &  82.54 & 1.41 & 6.01  & 6.33        \\
6. & HD 51608 c  &       18    &        3.0-4.0   & 0.7-1.3 & 0.95-1.10 &     95.4  &       0.379  &  82.99 & 1.37 & 6.13  & 6.44        \\
7. & Gl 785 b    &       21.6  &        3.0-4.0   & 0.7-1.3 & 0.95-1.10 &     74.4  &       0.319  &  77.42 & 1.66 & 5.97  & 6.42    \\
8. & HD 104067 b &       59.1  &        5.0-6.0   & 0.7-1.5 & 1.25-1.40 &     55.8  &       0.264  &  84.92 & 1.43 & 4.97  & 5.38    \\
9. & HD 3651 b   &       60    &        5.0-6.0   & 0.7-1.5 & 1.25-1.40 &     62.2  &       0.284  &  86.85 & 1.34 & 4.98  & 5.39    \\
10. & HD 16141 b  &       68.3  &        5.0-6.0   & 0.7-1.5 & 1.25-1.40 &     75.8  &       0.350  &  87.66 & 1.27 & 4.84  & 5.24    \\
\enddata
\tablecomments{The planetary radii were estimated from the masses. The meaning of other parameters is the same as in Table \ref{syspar} except that T$_{50}$ was determined for 10 transits only.}                              
\end{deluxetable}

\end{document}